\documentclass[prb,twocolumn,showpacs,preprintnumbers,amsmath,amssymb]{revtex4}
\usepackage{amsfonts}
\usepackage{latexsym}
\usepackage{graphicx}
\usepackage{color}

\begin{document}

\title{Kinetics and Boltzmann kinetic equation for fluctuation Cooper
pairs}

\author{Todor M. Mishonov}\email[E-mail address: ]{todor.mishonov@fys.kuleuven.ac.be}
\author{Georgi V. Pachov}
\author{Ivan N. Genchev}
\author{Liliya A. Atanasova}
\author{Damian Ch. Damianov}

\affiliation{Laboratorium voor Vaste-Stoffysica en Magnetisme,
Katholieke Universiteit Leuven, Celestijnenlaan 200 D, B-3001
Leuven, Belgium}
\affiliation{Department of Theoretical Physics,
Faculty of Physics, Sofia University St.~Kliment Ohridski, 5
J.~Bourchier Blvd., Bg-1164 Sofia, Bulgaria}

   \preprint{cond.tex \today}
\begin{abstract}
The Boltzmann equation for excess Cooper pairs above $T_c$ is
derived in the framework of the time-dependent Ginzburg-Landau
(TDGL) theory using Langevin's approach of the stochastic
differential equation. The Newton dynamic equation for the
momentum-dependent drift velocity is obtained and the effective
drag force is determined by the energy dependent life time of the
metastable Cooper pairs. The Newton equation gives just the Drude
mobility for the fixed momentum of Cooper pairs. It is shown that
the comparison with the well-known result for Aslamazov-Larkin
paraconductivity and BCS treatment of the excess Hall effect can
give the final determination of all the coefficients of TDGL
theory. As a result the intuitive arguments used for an
interpretation of the experimental data for fluctuation kinetics
are successively introduced. The presented simple picture of the
degenerated Bose gas in $\tau$-approximation near the
Bose-Einstein condensation temperature can be used for analysis of
fluctuation conductivity for the cases of high frequency and
external magnetic field for layered and bulk superconductors. The
work of the Boltzmann equation is illustrated by
frequency-dependent Aslamazov-Larkin conductivity in nanowires, in
the two-dimensional case and in the case of strong electric field
where the TDGL equation is solved directly. There are also derived
explicit formulas for the current in the case of arbitrary time
dependence of electric field up to THz range, the distribution of
fluctuation Cooper pairs for nonparabolic dispersion, the
influence of the energy cut-off and the self-consistent equation
for the reduced temperature. The general theory is illustrated by
formulas for fluctuation conductivity in nanowires and
nanostructured superconductors.
\end{abstract}

\maketitle

\section{Introduction}
For all high temperature superconductors the fluctuation phenomena
can be observed and their investigation takes a significant part
of the complete understanding of these materials; for a
contemporary review on the fluctuation phenomena in
superconductors see the review by Larkin and
Varlamov\cite{Larkin02}. The Ginzburg-Landau (GL) approach of the
order parameter is an adequate tool to investigate the
low-frequency behavior of fluctuations near to $T_c;$ for a review
of the Gaussian GL fluctuations see Ref.~\onlinecite{Mishonov00}.
A lot of important papers on the fluctuation phenomena in
superconductors and related topics have not been cited in these
reviews, see for example Ref.~\onlinecite{Ausloos}. We wish to
point out that the GL approach is the standard tool for the
investigation of magnetic field penetration in
superconductors\cite{Vos:01} and even non-Gaussian approach to
critical fluctuations.\cite{Tuszynski:91}

Amidst all kinetic phenomena the fluctuation conductivity created
by the metastable in the normal state Cooper pairs is probably
best investigated. The Boltzmann equation is a standard tool for
investigation of kinetic phenomena and the purpose of the present
paper is to derive the Boltzmann equation for fluctuation Cooper
pairs and to illustrate its work on the example of the fluctuation
conductivity; a shortened version of the present research was
presented in preliminary communications.\cite{Mishonov-Damianov}
We rederive the frequency dependence of the Aslamazov-Larkin
conductivity, fluctuation Hall effect at weak magnetic fields, and
magnetoconductivity. We analyze the experimental data for indium
oxide films and find significant deviation from the BCS weak
coupling prediction. We are coming to the conclusion that a
systematic investigation of lifetime of fluctuation Cooper pairs
will give important information for our understanding of the
physics of superconductivity.

\section{From TDGL equation via Boltzmann equation to Newton equation}
%
Our starting point is the time-dependent Ginzburg Landau (TDGL)
equation for the superconducting order parameter derived in the
classical paper by Abrahams and Tsuneto,\cite{Abrahams:66} see
also Ref.~\onlinecite{Abrahams:68} and references cited in the
review by Larkin and Varlamov\cite{Larkin02}
\begin{equation}\label{TDGL}
\frac{\left(-i\hbar D_r\right)^2}{2m^*}\Psi +a\Psi+b|\Psi|^2\Psi=
- \hbar\gamma \left(D_t\Psi-\zeta\right),
\end{equation}
where $m^*$ and $|e^*|=2|e|$ are the mass and charge of the Cooper
pairs, parameter $\gamma$ describes the dissipation, and
$\zeta(\mathbf r,t)$ is the external noise in TDGL equation. Here
\begin{eqnarray}
-i\hbar D_r &=& -i\hbar\nabla -e^*\mathbf{A}/c,\nonumber\\
 i\hbar D_t &=&  i\hbar\partial_t -e^*\varphi,\nonumber
\end{eqnarray}
are the operators of kinetic momentum and energy, $\mathbf{A}$ is
a vector-potential, and $\varphi$ is the potential.

Close to the critical temperature $a(T)\approx (T-T_c)a_0/T_c,$
and $b\approx \mathrm{const},$ where $a_0=\hbar^2/2m^*\xi^2(0),$
and $\xi(0)$ is coherence length.

The correlations of the white noise  $\langle\zeta\rangle=0,$
\begin{equation}
\langle\,\zeta^*(\mathbf r_1,t_1)\;\zeta(\mathbf r_2,t_2)\,
\rangle =\Gamma\,\delta(t_1-t_2)\,\delta(\mathbf{r}_1-\mathbf
r_2),
\end{equation}
are parameterized by  fluctuation parameter $\Gamma.$ The BCS
theory gives $$\gamma_{_{BCS}}=\frac{\pi}{8}\frac{a_0}{T_c},$$ and
that is why we parameterize $\gamma=\gamma_{_{BCS}}\tau_{rel}, $
by the dimensionless parameter $\tau_{rel}\simeq 1,$ which
describes the relative  life-time of fluctuation Cooper pairs.

The most simple is the case of free particle, which means $
\mathbf A=0 $, \,$ \varphi=0 $, \,$ b|\Psi|^2\approx 0.$
Introducing the  Fourier transformation
\begin{align}
\Psi(\mathbf r,t)&=\sum_p\frac{e^{i\mathbf p\cdot\mathbf r/\hbar}}
{\sqrt\mathcal{V}}\,\psi_p(t),  \\[7pt] 
\zeta(\mathbf r,t)&=\sum_p\frac{e^{i\mathbf p\cdot\mathbf
r/\hbar}} {\sqrt\mathcal{V}}\,\zeta_p(t),
\end{align}
where
$$\sum_p\approx\mathcal{V}\int\frac{d^D\,p}{(2\pi\hbar)^D},$$
and $$\langle\,\zeta_p^*(t_1)\;\zeta_q(t_2)\,\rangle=\Gamma
\delta_{p,q}\,\delta(t_1-t_2),$$
we obtain TDGL equation in momentum representation
\begin{equation}
(\varepsilon_p+a)\psi_p=-\hbar\gamma(d_t\psi_p-\zeta_p).
\end{equation}
The solution of this reads
\begin{equation}
\psi_p(t)=e^{-t/2\tau_p}\left(\int_0^t
e^{t'/2\tau_p}\,\zeta_p(t')dt'+\psi_p(0)\right),
\end{equation}
where
\begin{equation}\label{tau}
\tau_p=\frac{\hbar\gamma}{2(\varepsilon_p+a(T))}, \qquad
\varepsilon_p=\frac{p^2}{2m^*}
\end{equation}
are momentum-dependent lifetime and kinetic energy of fluctuation
Cooper pairs. The number of particles for every momentum  can be
found by noise averaging
\begin{equation}
n_p=\langle\,\psi_p^*(t)\psi_p(t)\rangle=
n_p(0)e^{-t/\tau_p}+(1-e^{-t/\tau_p})\bar{n}_p,
\end{equation}
where $n_p(0)=|\psi_p(0)|^2$ is the initial number. The time
differentiation of this solution gives the well-known Boltzmann
equation
\begin{equation}
\frac{d}{dt}n_p(t)=-\frac{1}{\tau_p}(n_p(t)-\bar n_p),
\end{equation}
which can be considered in this physical situation as a
consequence of the TDGL equation. The quantity
$$ \bar n_p=n_p(t=\infty)=\Gamma\tau_p$$
gives the equilibrium number of  particles. The fluctuation
parameter $\Gamma$ is related to dissipation parameter $\gamma$ by
the fluctuation-dissipation theorem, which here takes the form
\begin{equation}
\Gamma=\frac{2T}{\hbar\gamma}=\frac{\bar
n_p}{\tau_p}=\frac{T}{a_0\tau_0},
\end{equation}
where $$ \bar n_p=\frac{T}{\varepsilon_p+a(T)}$$
is the Rayleigh-Jeans distribution.

Let us now analyze the influence of a weak electric field in the
Boltzmann\cite{Boltzmann:12} equation
\begin{equation}
\label{Boltzmann}
\partial_t n_p+ e^*\mathbf{E}\cdot\partial_{\mathbf p}
n_p=-\frac{1}{\tau_p}(n_p-\bar{n}_p).
\end{equation}

Quantum mechanics was born during Halle conference in 1891, when
exposed to the ignorant criticism of both statistical methods and
atomic physics, Boltzmann suddenly made a remark: "I see no reason
why energy shouldn't also be regarded as divided atomically"
Ref.~\onlinecite{Flamm:97}. Later on applying the Boltzmann method
to the problem of black body radiation Planck found that the
constant appearing in the photon spectrum is just the volume of
the Boltzmann cells in the phase space. Due to this reason, Planck
called the quantity $2\pi\hbar$ after Boltzmann - Boltzmann
constant.

For the solution we search in the form
\begin{equation}
n_p(t)=n(\mathbf p,t)\approx\bar n\left(\mathbf p- m^*\mathbf
V(\mathbf p,t)\right),
\end{equation}
and we obtain the Newton equation
\begin{equation}
m^*d_t\mathbf V_p(t)=e^*\mathbf E-\frac{m^*}{\tau_p}\mathbf V_p(t)
\end{equation}
for the field of drift velocity in momentum space. The general
formula for the current gives
\begin{equation}
\label{current} \mathbf j_{\,\mathrm{fl}}=\sum_p
e^*\frac{n_p}{\mathcal V}\mathbf
v_p=\tensor{\sigma}_{\mathrm{fl}}\cdot\mathbf E, \qquad
n_{_{\mathrm D}}=\sum_p\frac{n_p}{\mathcal V},
\end{equation}
where $n_D$ is the $D$-dimensional volume density of the
fluctuation Cooper pairs. Substitution here of the shifted
equilibrium distribution gives the well-known formula for the
conductivity tensor\cite{LL:10}
\begin{equation}
\label{sigmatensor}
\tensor\sigma_\mathrm{fl}=e^{*2}\int\frac{d^Dp}{(2\pi\hbar)^D}
\frac{\mathbf v_p\otimes\mathbf v_p}{1/\tau_p-i\omega}
\left(-\frac{\partial\, n_p}{\partial\, \varepsilon_p}\right),
\end{equation}
where $ \mathbf v_p=\partial_p\varepsilon_p=\mathbf p/m^*$ is the
Cooper pairs' velocity. This is only a small fraction of the total
conductivity
\begin{equation}
\sigma(T)=\sigma_{_\mathrm N}(T)+\sigma_\mathrm{fl}(\epsilon),
\quad \epsilon\equiv\ln\frac{T}{T_c}\approx\frac{T-T_c}{T_c},\quad
\sigma_\mathrm{fl}\ll\sigma_{_\mathrm N}.
\end{equation}
For thin superconducting films $D=2$ substituting
\begin{equation}
\frac{dp_xdp_y}{(2\pi\hbar)^2}= \frac{d(\pi p^2)}{(2\pi\hbar)^2}=
\frac{m^*}{2\pi\hbar^2}d\varepsilon_p, \quad -\frac{\partial\bar
n}{\partial\varepsilon}=\frac{T}{(\varepsilon+a)^2},
\end{equation}
we obtain the classical result by Aslamazov and
Larkin\cite{Aslamazov68}
\begin{equation}\label{AL68}
\sigma_{_\mathrm{AL}}(\epsilon)=\frac{e^2}{16\hbar}\tau_\mathrm{rel}\frac{T_c}{T-T_c}
=\frac{e^2T}{\pi\hbar^2}\tau(\epsilon),
\end{equation}
where
\begin{equation}
\tau(\epsilon) \equiv \tau(\mathbf
p=0,\epsilon)=\frac{\pi\hbar}{16T_c}
\frac{\tau_\mathrm{rel}}{\epsilon}=\frac{\tau_0}{\epsilon}
\end{equation}
is the lifetime for Cooper pairs with zero momentum.

For the two-dimensional (2D) case conductivity is just the inverse
resistance $\sigma^{(2D)}(T)=R_\square^{-1}(T).$  For conventional
disordered superconductors normal conductivity can be approximated
by residual conductivity far above $T_c,$ for example, $T=3T_c.$
In this approximation Aslamazov-Larkin conductivity can be
rewritten in a convenient for experimental data processing form
\begin{equation}
\left(\frac{1}{R_\square(T)}-\frac{1}{R_\square(3T_c)}\right)^{-1}\approx
\frac{16\hbar}{e^2\tau_\mathrm{rel}}\left(\frac{T}{T_c}-1\right).
\end{equation}

Performing the linear regression fit of the data presented in
Ref.~\onlinecite{Fiory83} we have obtained that for indium oxide
films $\,\tau_\mathrm{rel}=1.15.$ This significant 15~\% deviation
from the weak coupling BCS value is created by strong coupling
effects. We conclude that analogous systematic investigations for
thin films would be very helpful for our understanding of the
dynamics of the order parameter in superconductors. Decreasing the
lifetime and $\tau_\mathrm{rel}$ by depairing impurities or
disorder for anisotropic gap superconductors definitely deserves a
great attention.

\section{Fluctuation conductivity in different physical condition}
\subsection{High frequency conductivity}

For diagonal components of conductivity taking into account that
$\mathrm{Tr}\openone= D$, from the general formula
Eq.~(\ref{sigmatensor}), we obtain
\begin{equation}
\label{sigmascalar}
\sigma_\mathrm{fl}=\frac{e^{*2}}{D}\int\frac{d^Dp}{(2\pi\hbar)^D}
\frac{v_p^2}{1/\tau_p-i\omega} \left(-\frac{\partial\,
n_p}{\partial\, \varepsilon_p}\right).
\end{equation}
It is convenient to introduce a dimensionless frequency
$z=\omega\tau(\epsilon)$. In order to derive the dimensionless
complex conductivity $\varsigma(\omega)$ we need to solve the
elementary integral
\begin{eqnarray}
   \varsigma(z) &=& \varsigma_1(z) + i\varsigma_2(z) \nonumber\\[7pt]
            &=&  2\int_1^\infty \frac{x-1}{x^2(x+y)}dx \label{integral}\\[7pt]
                &=& \frac{2}{y}\left[ \left(
                     1+\frac{1}{y}\right) \ln(1+y) -1 \right],
                    \label{Matsubara}\\[7pt]
            &=& \frac{2}{-iz}\left[ \left(
                     1+\frac{1}{-iz}\right) \ln(1-iz) -1
                     \right],\nonumber
\end{eqnarray}
where $x=p^2/2m^*a(\epsilon)+1$ is the kinetic energy of Cooper
pairs taken into account from the ``chemical potential'' in
$a(\epsilon)$ units, and $y=-iz=-i\omega\tau(\varepsilon)$ is the
dimensionless Matsubara frequency $\zeta_{_\mathrm M}=-i\omega.$
The integral Eq.~(\ref{integral}) is solved considering the
Matsubara frequency $y$ to be a real variable. Then we can make
the analytical continuation to real frequencies substituting
$y=-iz$ in the result Eq.~(\ref{Matsubara}). This method is very
popular in the quantum field theory, but works effectively for
classical problems as well. In such a way we obtain
\begin{eqnarray}
   \varsigma_1(z) &=& \frac{2}{z^2}\left[z\arctan(z) -
                      \frac{1}{2}\ln(1+z^2) \right]
      \nonumber\\ &=& \frac{2}{\pi}\mathcal{P}\int_0^\infty
                     \frac{x\varsigma_2(x)}{x^2-z^2}dx,\\
   \varsigma_2(z) &=& \frac{2}{z^2}\left[\arctan(z) -z
                      +\frac{z}{2}\ln(1+z^2) \right]
      \nonumber\\ &=& -\frac{2z}{\pi}\mathcal{P}\int_0^\infty
                     \frac{\varsigma_1(x)}{x^2-z^2}dx.
\end{eqnarray}
Then the frequency-dependent conductivity reads
\begin{equation}
   \sigma_{_\mathrm{2D}}(\epsilon,\omega)
    =\sigma_{_\mathrm{AL}}(\epsilon)\,
    \varsigma(\omega \tau(\epsilon)).
\end{equation}
The integral Eq.~(\ref{sigmascalar}) can be solved for arbitrary
dimension
\begin{equation}
   \sigma_{_\mathrm{D}}(\epsilon,\omega)
        =\sigma_{_\mathrm{D}}(\epsilon)\, \varsigma_{_\mathrm{D}}(z),\qquad
\end{equation}
cf. the paper by Dorsey\cite{Dorsey}
\begin{equation}
   \sigma_{_\mathrm{D}}(\epsilon) = 4\frac{\Gamma(2-D/2)}{(4\pi)^{D/2}}
    \frac{e^2}{\hbar}\left[\xi(\epsilon)\right]^{2-D}
    \frac{T\tau(\epsilon)}{\hbar},
\end{equation}
where $\xi(\epsilon) \equiv \xi(0)/\sqrt{\epsilon}$
is the temperature-dependent coherence length. The conductivity in
this case has the form
\begin{eqnarray}
   \varsigma_{_{1,\mathrm D}}(z) &=& \frac{8}{D(D-2)z^2}
   \left[1\right.\nonumber\\
&&\left.\quad -(1+z^2)^{D/4}\cos\left(\frac{D}{2}\arctan z
\right)\right] ,\nonumber\\
   \varsigma_{_{2,\mathrm D}}(z) &=& \frac{8}{D(D-2)z^2}
\left[-\frac{D}{2}z\right.\nonumber\\ &&\left.\quad
+(1+z^2)^{D/4}\sin\left(\frac{D}{2}\arctan z \right)\right].
\end{eqnarray}
%

\subsection{Hall effect}

The fluctuation Hall conductivity also can be derived in the
framework of the Boltzmann kinetic equation. We have to take into
account a small imaginary part $\alpha$ of $\gamma$ parameter in
the TDGL equation, i.e., $\gamma\rightarrow \gamma+i\alpha,$  and
$\alpha\ll\gamma.$ The solution\cite{Mishonov-Damianov} of the
kinetic equation gives
\begin{equation}
 \sigma_{xy}(\epsilon)=\frac{Z}{3}\omega_c\tau(\epsilon)
 \sigma_{\mathrm{AL}}(\epsilon) \propto \tau^2(\epsilon),
\end{equation}
where $\omega_c=e^*B/m^*c$ is the ``cyclotron'' frequency and
\begin{equation}
Z=-\textrm{Im}\frac{1}{\gamma+i\alpha} \approx
\frac{\alpha}{\gamma^2} \ll 1.
\end{equation}
This result agrees with microscopic calculations.\cite{Larkin02}
Due to the small value of the parameter $\alpha$, fluctuation Hall
effect is difficult to observe. With fitting of $\alpha$ and $m^*$
from the experimental data finishes the complete determination of
parameters of TDGL theory.

\subsection{Magnetoconductivity}

It is interesting to mention that the classical formula for the
conductivity Eq.~(\ref{sigmatensor}) correctly works even for
strong magnetic fields. We only have to substitute the momentum
integration with summation on discrete Landau levels, taking into
account the density of Landau magnetic subbands
\begin{equation}
 \epsilon_p\rightarrow\epsilon_n=\hbar\,\omega_c\left(n+\frac{1}{2}\right)=
 a_0(2n+1)h,
\end{equation}
where
\begin{equation}
 h=\frac{\hbar\,\omega_c}{2a_0}=\frac{B_z}{B_{\mathrm{c2}}(0)}
\end{equation}
is the dimensionless magnetic field and
\begin{equation}
 B_{\mathrm{c2}}(0)=-T_c\frac{d}{dT} B_{\mathrm{c2}}(T)|_{T_c}
\end{equation}
is the linear extrapolation. In the numerator of
Eq.~(\ref{sigmatensor}) we have to substitute the classical
velocity with the oscillator matrix elements of the momentum.
Analogously for the energy-dependent lifetime we have to average
on neighboring levels. Due to the triviality of the oscillator
problem these substitutions can be performed in only one way, and
Aslamazov-Larkin conductivity Eq.~(\ref{AL68}) is substituted by
the magnetoconductivity of Abrahams, Prange and Stephen\cite{APS}
(APS)
\begin{equation}
 \sigma_{_{\mathrm{AL}}}(\epsilon)=\frac{e^2T\tau_0}{\pi\hbar}\,
 \frac{1}{\epsilon}\rightarrow
 \sigma_{_{\mathrm{APS}}}(\epsilon,h)=\frac{e^2T\tau_0}{\pi\hbar}
 f(\epsilon,h),
\end{equation}
i.e., $1/\epsilon$ has to be substituted by APS function
\begin{equation}
 f(\epsilon,h)=
 \frac{2}{h^2}\left[ \epsilon \digamma\left(\frac{1}{2}+
 \frac{\epsilon}{2h}\right)-
 \epsilon\digamma\left(1+\frac{\epsilon}{2h}\right)+h
 \right].
\end{equation}
This two-dimensional result can be easily generalized for layered
and bulk superconductors using the layering operator introduced in
Ref.~\onlinecite{Mishonov00}.

\subsection{Strong electric fields}

Using the optical gauge $$ \varphi = 0,\qquad
\mathbf{A}=-t\mathbf{E}, $$ the TDGL equation Eq.~(\ref{TDGL})
reads
\begin{equation}
d_u\psi_q(u) = -\frac{1}{2}\left[(q+fu)^2 +
\epsilon\right]\psi_q(u) + \bar{\zeta}_q(u),
\end{equation}
where we are introducing dimensionless variables for the
$u=t/\tau_0$ time, $q=p\xi(0)/\hbar$  momentum,
$f=e^*E\tau_0\xi(0)/\hbar$ electric field, and
$\bar{\zeta}_q(u)=\tau_o\zeta_p(t)$ noise. We have a linear
ordinary differential equation which can be solved for arbitrary
$f(u),$ i.e., for arbitrary time dependence of the electric field.
For constant electric field the TDGL equation has the solution
\begin{eqnarray}
\psi_q(u) &=& \left\{\int_0^u
\exp\left[\frac{1}{2}\int_0^{u_1}\left[(q+fu_2)^2 +
\epsilon\right]du_2\right]\right.\nonumber\\[7pt]
&&\qquad\qquad\quad\times \left.\bar{\zeta_q}(u_1)du_1+
\psi_q(0)\right\}\nonumber\\[7pt] &&\times
\exp\left[-\frac{1}{2}\int_0^u\left[(q+fu_3)^2 +
\epsilon\right]du_3\right].
\end{eqnarray}
In order to obtain the static [$t\gg \tau(\epsilon)$] momentum
distribution we have to perform the noise averaging
\begin{eqnarray}
\label{distribution}
 n_k &=& \lim_{u\rightarrow\infty}
  \langle \left|\psi_{q+fu}(u)\right|^2\rangle\\[7pt]
    &=& \frac{T}{a_0}\int_0^\infty \exp\left[-\left(k^2+\epsilon\right)v +
  fkv^2 -\frac{1}{3}f^2v^3\right]dv,\nonumber
\end{eqnarray}
where $u_1=u-v$ and $$
k=q+fu=\left(p-e^*A\right)\frac{\xi(0)}{\hbar} $$ is the
dimensionless kinetic momentum. This distribution can be directly
derived from Boltzmann equation Eq.~(\ref{Boltzmann}) for
fluctuation Cooper pairs.\cite{Mishonov02} In
Ref.~\onlinecite{Mishonov02} it was demonstrated that substitution
of the momentum distribution Eq.~(\ref{distribution}) in the
formula for the current density Eq.~(\ref{current}) gives the
result which agrees with the formula by Dorsey\cite{Dorsey}; cf.
also the paper by Gor'kov\cite{Gor'kov}
\begin{equation}
j(E_x)=\frac{e^2\tau_\mathrm{rel}E_x}{16\hbar\left[2\pi^{1/2}\xi(0)\right]^{D-2}}
\int_0^\infty\frac{\exp\left(-\epsilon u-gu^3\right)}{u^{(D-2)/2}}
\,d u, \label{currentD}
\end{equation}
where
$$
g \equiv \frac{f^2}{12}, \quad f=\frac{e^*E_x\xi(0)\tau_0}{\hbar}
=\frac{\pi}{8}\,\frac{eE_x\xi(0)}{T_c}\,\tau_\mathrm{rel}.
$$
Differentiating the upper expression we obtain differential
conductivity
\begin{eqnarray}
    \sigma_{\mathrm{diff}}&=&\frac{dj(E_x)}{dE_x}=
    \frac{e^2\tau_{\mathrm{rel}}}{16\hbar(2\sqrt{\pi}\xi(0))^{D-2}}\nonumber\\[7pt]
    &&\times\int_0^\infty \frac{1-2gu^3}{u^{(D-2)/2}}\exp(-\epsilon
    u-gu^3)\,du.
\end{eqnarray}
Applying a voltage
$U(t)=U_{_\mathrm{DC}}+U_{_\mathrm{AC}}\cos\omega t$ to the
nanowire, the differential conductivity can be easily determined
measuring the AC component for the current if $U_{_\mathrm{AC}}\ll
U_{_\mathrm{DC}}$. Cooling the sample the differential
conductivity will decrease, then at some temperature it will be
annulated and what will happen at further cooling is an
interesting experimental question.

\section{Current functional: self-consistent approximation and energy cut-off}

The self-consistent approximation for the reduced critical
temperature\cite{Mishonov00,Dorsey} in the one-dimensional (1D)
case reads
\begin{equation}
\label{renormalizedT}
  \epsilon_{\mathrm{ren}}= \ln \frac{T}{T_0}+\frac{b}{a_0}n_{_{1D}}=
 \ln \frac{T}{T_0}+\epsilon_{_{1G}}{\cal N}_1(\epsilon_{\mathrm{ren}},f),
\end{equation}
where $n_{_{1D}}$ is the bulk density of the fluctuation Cooper
pairs when we have 1D fluctuations in a wire with cross section $S
\ll \xi^2(\epsilon)$ and
\begin{equation}
\epsilon_{_{1G}}\equiv\frac{\mu_0\lambda^2(0)\xi(0)e^2T_c}{\sqrt{\pi}S\hbar^2}
 =\frac{k_B}{8\sqrt{\pi}\Delta C\xi(0)S},
 \label{Gi1D}
\end{equation}
where $\lambda(\epsilon)=\lambda(0)/{\sqrt{-\epsilon}}$ is the
temperature-dependent penetration depth and $\Delta C$ is the jump
of the specific heat at $T_c$ per unit volume. For numerical
calculations the function
\begin{equation}
\label{density1D}
 {\cal N}_1(\epsilon,f)\equiv\int_0^\infty
 \exp(-\epsilon v-gv^3)\frac{dv}{\sqrt{v}}
\end{equation}
has to be programmed as
\begin{equation}
{\cal N}_1(\epsilon,f)=2\int_0^\infty
 \exp(-gz^6-\epsilon z^2)dz, \quad z^2=v.
\end{equation}
Analogously for the thin superconducting film with thickness $d_f
\ll \xi(\epsilon)$ the equation for reduced temperature at zero
electric field takes the form\cite{Mishonov00}
\begin{equation}
  \epsilon_{\mathrm{ren}} = \ln \frac{T}{T_0}+\frac{b}{a_0}n_{_{2D}}
  =\ln \frac{T}{T_0}+\epsilon_{_{2G}}{\cal N}_2(\epsilon_{\mathrm{ren}}),
\end{equation}
where $n_{_{2D}}$ is the volume density of the fluctuation Cooper
pairs having 2D fluctuations,
\begin{equation}
\epsilon_{_{2G}} \equiv \frac{k_B}{4\pi \Delta C\xi^2(0)d_f} =2\pi
\mu_0\frac{T_c}{d_f}\left(\frac{\lambda(0)}{\Phi_0}\right)^2
\end{equation}
is the 2D Ginzburg number and
\begin{equation}
{\cal N}_2(\epsilon)
 \equiv\ln\left(\frac{c+\epsilon}{\epsilon}\right).
\end{equation}

As simplest possible application of these results we have to
mention nanostructured superconductors, e.g.,
nanowires\cite{nanowires}, similar to those used for long time for
investigation of phase slip centers\cite{phase-slip_centers} of
the superconducting phase. We are pointing out that
paraconductivity is a property of the normal phase.

For general (nonparabolic) dispersion we can derive from the TDGL
equation the formula for the distribution of fluctuation Cooper
pairs
\begin{eqnarray}
  n_k(u) &=& \frac{T}{a_0}\int_0^u \exp\left\{-\int_{u_1}^u\left[\frac{\varepsilon(k(u_2))}{a_0}
  +\epsilon\right]du_2\right\}du_1  \nonumber\\[7pt]
  &&
  \!+\,\bar{n}_k(0)\exp\left\{-\int_{u_1}^u
  \left[\frac{\varepsilon(k(u_2))}{a_0}+\epsilon\right]du_2\right\}\!,
\end{eqnarray}
where the dimensionless kinetic momentum and the vector potential
are
$$k(u)=q+\overline{A}(u),\quad \overline{A}(u)=-\frac{e^*\xi(0)}{\hbar}A(t).$$
In the case of parabolic dispersion and arbitrary time dependence
of the electric field we can write Eq.~(\ref{current}) for the
current functional in the form
\begin{equation}\label{current[A]}
 j[A]=\frac{\sqrt{\pi}\hbar e}{m^*\xi(0)}\left[\frac{T}{a_0}\int_0^u
 F_A[u_1]du_1+\bar{n}_k(0)F_A[0]\right],
\end{equation}
where for brevity we introduce the functionals
\begin{eqnarray}\label{FA}
 F_A[u_1]& \equiv & \left(\frac{\overline{A}(u)}{\sqrt{u-u_1}} -
       \frac{B_A[u_1]}{(u-u_1)^{3/2}}\right)\\[7pt]
  &&\times\exp\left[\frac{(B_A[u_1])^2}{u-u_1}-G(u_1)\right]
\end{eqnarray}
and
$$B_A[u_1] \equiv \int_{u_1}^u \overline{A}(u_2)du_2,$$
$$G_A[u_1] \equiv -\epsilon(u-u_1)+\int_{u_1}^u(\overline{A}(u_2))^2du_2.$$

Local GL theory with an energy cut-off for the kinetic energy
$\varepsilon_p<a_0c$ [or for the kinetic energy taken into account
from the chemical potential $\varepsilon_p+a(T)<a_0c$] is very
often used for describing the fluctuation phenomena
\cite{Mishonov00}
\begin{equation}\label{ep}
    \varepsilon(k)=\left\{\ \begin{array}{cc}
      a_0k^2,           & \quad|k\,|<\Lambda \\[7pt]
      a_0c,           & \quad|k\,|>\Lambda\\
    \end{array}
    \right.
\end{equation}
where the dimensionless constant $\Lambda\equiv\sqrt{c}\simeq 1.$
For YBa$_2$Cu$_3$O$_{7-\delta}$ the recent
investigations\cite{Silva:03} of high-frequency fluctuation
conductivity determined $\Lambda\approx 0.5.$ See also the recent
work by Puica and Lang, Ref.~\onlinecite{Puica}, where the effects
of energy cut-off and self-consistent interaction are treated in
great details.
Then the functional which participates in the formula for the
current Eq.~(\ref{current[A]}) takes the form
\begin{eqnarray}
  &&F_A[u_1]=-\frac{\sinh(2\Lambda B_A[u_1])}{\sqrt{u-u_1}}\nonumber\\[7pt]
   &&\qquad\times\exp\left[-(u-u_1)
   \left(c+\frac{(B_A[u_1])^2}{(u-u_1)^2}\right)\right]\\[7pt]
  &&\quad+\left(\overline{A}(u)-\frac{B_A[u_1]}{(u-u_1)}\right)
     \exp\left(\frac{(B_A[u_1])^2}{u-u_1}-G_A[u_1]\right)\nonumber\\[7pt]
  &&\qquad\times\int_{-\Lambda-A(u)}^{\Lambda-A(u)}
     \exp\left[-(u-u_1)\left(q+\frac{B_A[u_1]}{u-u_1}\right)^2dq\right].\nonumber
\end{eqnarray}
These rather complicated formulas are necessary for investigation
of paraconductivity in the THz range. In the next section we will
give an illustration for the important one dimensional case.

\section{Fluctuation conductivity in nanowires}

The recent development of the technology of the performance of
nanowires made it possible and even indispensable for the
investigation of fluctuation conductivity. In this section we will
analyze in detail the general results in 1D case.

The integrants in the momentum distribution
Eq.~(\ref{distribution}) is actually age distribution for
fluctuation Cooper pairs
\begin{equation}
{\cal F}(v;\,k,\epsilon,f)= \exp\left[-\left(k^2+\epsilon\right)v
 + fkv^2 -\frac{1}{3}f^2v^3\right],
\end{equation}
the variable $v$ is the age in units $\tau(\epsilon).$ Time
integration returns us to the momentum distribution, which using
the dimensional variables
\begin{equation}
k_\epsilon\equiv\frac{k}{|\epsilon|^{1/2}},\qquad
 f_\epsilon\equiv\frac{f}{|\epsilon|^{3/2}}
\end{equation}
reads
\begin{equation}
\overline{n}(k;\, \epsilon,f)=\frac{n_T}{|\epsilon|}
 {\cal F_\pm}(k_\epsilon,f_\epsilon),
\end{equation}
where
\begin{equation}
\label{fpm}
{\cal
F_\pm}(k_\epsilon,f_\epsilon)\equiv\!\int_0^\infty
 \exp\!\left[-(k_\epsilon^2\pm1)x+f_\epsilon k_\epsilon x^2
 -\frac{1}{3}f_\epsilon^2x^3
 \right]\! dx
\end{equation}
and $x=|\epsilon|u.$
\begin{figure}[htp]
\begin{center}
\includegraphics[width=.55\columnwidth,totalheight=0.25\textheight]{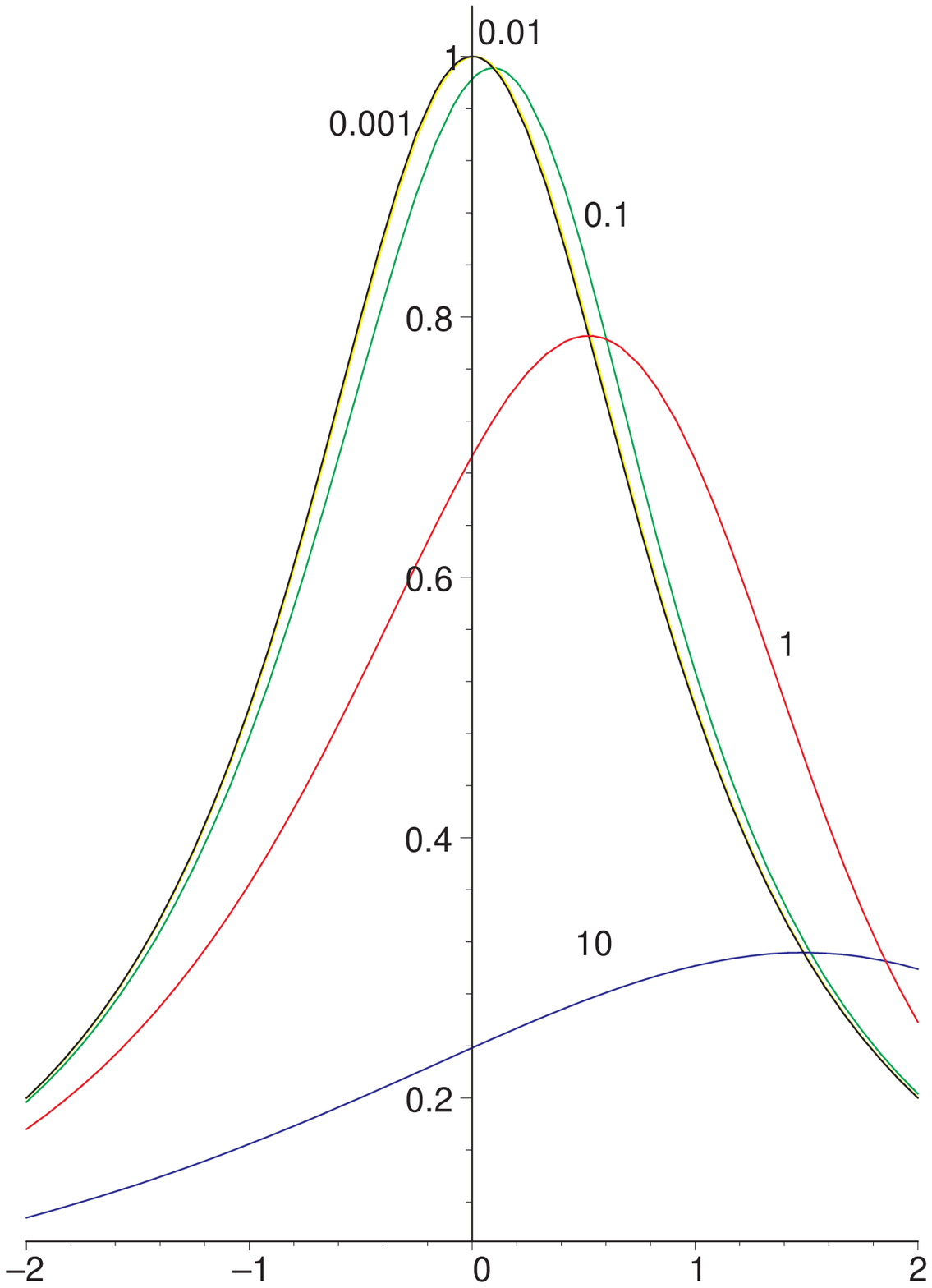}
\vspace*{0.2cm} \caption{{\footnotesize Age distribution for
fluctuation Cooper pairs above $T_c$ according to electric field
$f_\epsilon.$ At absciss is the momentum $ k_\epsilon $ and at
ordinate is the function ${\cal F_+}(k_\epsilon,f_\epsilon)$ from
Eq. (\ref{fpm}). The curves illustrate different values of
$f_\epsilon.$ The number of particles with zero kinetic momentum
in weak electric fields is much more than in strong fields. }}
\end{center}
\hfill
\begin{center}
\includegraphics[width=.55\columnwidth,totalheight=0.25\textheight]{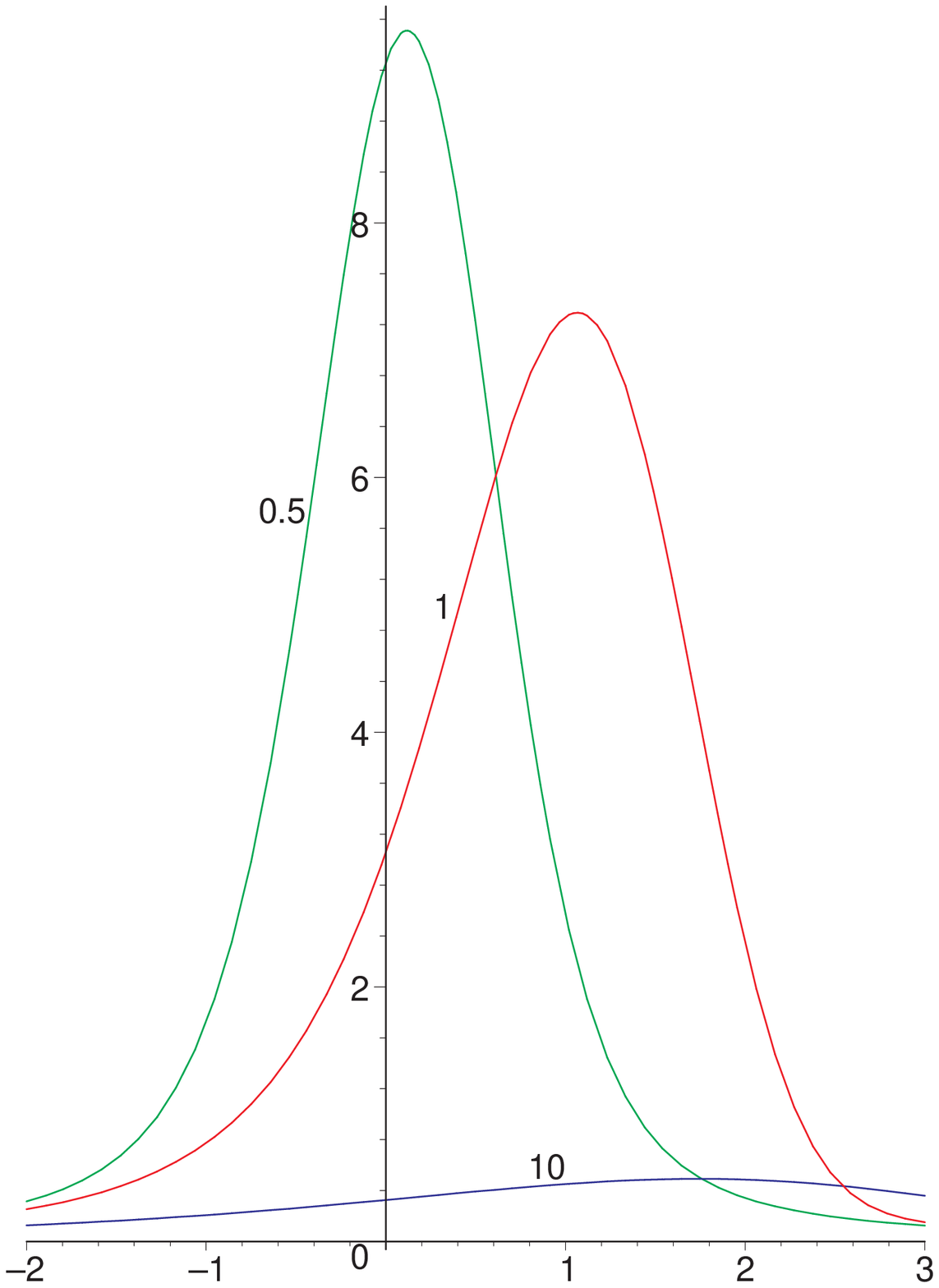}
\vspace*{0.2cm} \caption{{\footnotesize Age distribution for
fluctuation Cooper pairs under $T_c$ depending on electric field
$f_\epsilon.$ At absciss is the momentum $ k_\epsilon $ and at
ordinate is the function ${\cal F_-}(k_\epsilon,f_\epsilon)$ from
Eq. (\ref{fpm}). The curves illustrate different values of
$f_\epsilon.$ Decreasing the electric field the number of
particles increases. The maximum in strong electric fields is
right-shifted for particles with positive momentum. }
\vspace*{0.5cm}}
\end{center}
\end{figure}

For the case $\epsilon=0$, $T=T_c$ or when $f\rightarrow\infty$,
i.e.,
\begin{equation}
f_\epsilon=e^\ast E_x\xi(\epsilon)\tau(\epsilon)/\hbar\gg1,
 \qquad \xi(\epsilon)=\xi(0)/|\epsilon|^{1/2}
\end{equation}
we obtain
\begin{equation}
n(k,f)=\frac{n_T}{f^{2/3}}{\cal F}_0(k_f), \qquad
 k_f=\frac{k}{f^{1/3}},
\end{equation}
where in
\begin{equation}
\label{f0} {\cal F}_0(k_f)
 \equiv\int_0^\infty\exp\left[-k_f^2y+k_fy^2-\frac{1}{3}y^3
 \right]dy
\end{equation}
we use the transformation $y=f^{2/3}v.$
\begin{figure}
\begin{center}
\includegraphics[width=.55\columnwidth,totalheight=0.25\textheight]{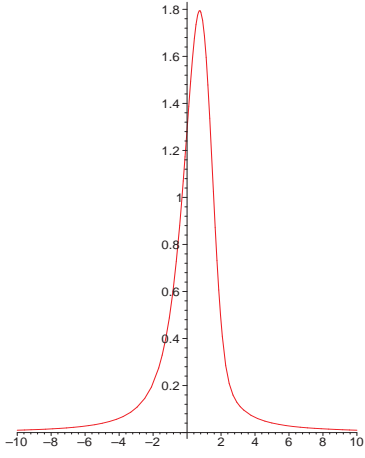}
\vspace*{0.2cm} \caption{{\footnotesize Momentum distribution for
fluctuation Cooper pairs in a strong electric field at $T_c.$ At
absciss is the scaling momentum $k_f$ and at ordinate is the
universal function ${\cal F}_0(k_f)$ from Eq. (\ref{f0}), which
describes particle distribution.}}
\end{center}
\end{figure}

For 1D case using Eq.~(\ref{currentD}), we express the current
\begin{equation}
j(E_x,\epsilon)
 =\frac{\sqrt{\pi}e^2}{8\hbar}\tau_\mathrm{rel}\xi(0)E_x
 {\cal J}(\epsilon,f)
\end{equation}
where
\begin{equation}
\label{currentf}
{\cal J}(\epsilon,f)
 =\int_0^\infty\exp(-\epsilon
 v-gv^3)\sqrt{v}dv, \quad
 g\equiv\frac{f^2}{12}.
\end{equation}
\begin{figure}
\hfill
\begin{center}
\includegraphics[width=.55\columnwidth,totalheight=0.25\textheight]{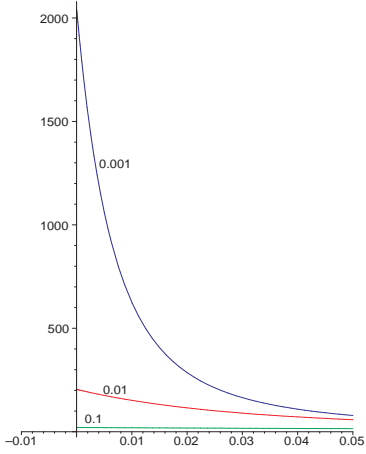}
\vspace*{0.2cm} \caption{{\footnotesize At ordinate is function
${\cal J}(\epsilon)$ participating in Eq.~(\ref{currentf}), which
describes the strength of the fluctuational current, and at
absciss is the dimensionless temperature $\epsilon$. The curves
show the dimensionless electric field $f.$ Coming nearer to
critical temperature $T_c$ the fluctuational current grows up
rapidly in very weak electric fields.}}
\end{center}
\end{figure}

The fluctuational current for Cooper pairs above and under the
critical temperature is
\begin{equation}
j=\frac{\pi e^2\tau_\mathrm{rel}\xi(0)E_x}{16\hbar|
 \epsilon|^{3/2}}\varsigma_{\pm} (f_\epsilon),
 \end{equation}
where
\begin{equation}
\label{sigmap}
\varsigma_\pm (f_\epsilon)
 =\frac{2}{\sqrt{\pi}}\int_0^\infty\exp(\mp
 v_\epsilon-g_\epsilon v_\epsilon^3)\sqrt{v_\epsilon}dv_\epsilon
\end{equation}
is the dimensionless function, which depends on the strength of
the electric field. For convenience we use
$g_\epsilon=f_\epsilon^2/12$ and $v_\epsilon=v|\epsilon|.$
We wish to point out the normalization and the asymptotics
\begin{equation}
\label{sigmas} \varsigma_+(0)=1, \qquad \varsigma_\pm
(f_\epsilon\rightarrow{\infty})\sim\frac{4}{\sqrt{3}f_\epsilon}.
\end{equation}
\begin{figure}
\begin{center}
\includegraphics[width=.55\columnwidth,totalheight=0.25\textheight]{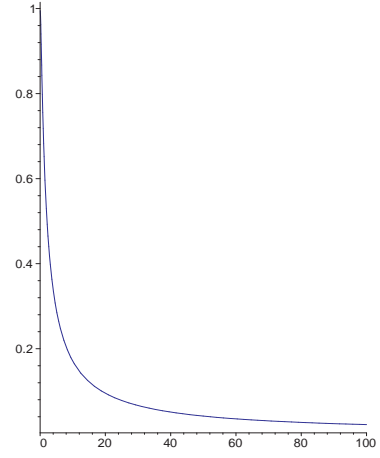}
\vspace*{0.2cm} \caption{{\footnotesize Distribution of
dimensionless function $\varsigma_+ (f_\epsilon)$ according to
Eq.~(\ref{sigmap}) above critical temperature $T_c.$ At absciss is
$f_\epsilon$. The maximum of fluctuational current is in zero
electric field near $T_c.$ }}\vspace*{0.5cm}
\end{center}
\end{figure}
\begin{figure}
\begin{center}
\includegraphics[width=.55\columnwidth,totalheight=0.25\textheight]{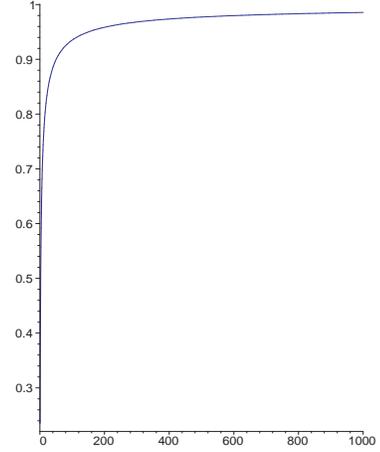}
\vspace*{0.5cm} \caption{{\footnotesize Asimptotic
$\frac{\sqrt{3}f_\epsilon}{4}\varsigma_+ (f_\epsilon)$ from
Eq.~(\ref{sigmas}) is going to constant when
$f_\epsilon\rightarrow{\infty}$. Above $T_c$ in very strong
electric field fluctuational current goes to constant. }}
\end{center}
\end{figure}
\begin{figure}
\begin{center}
\includegraphics[width=.55\columnwidth,totalheight=0.25\textheight]{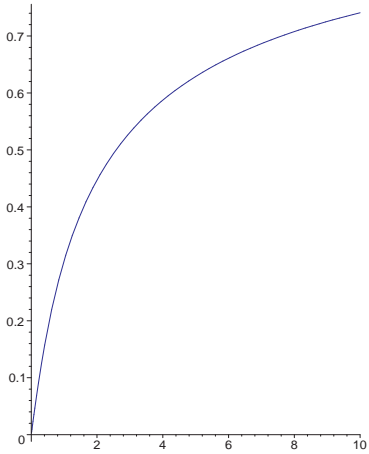}
\vspace*{0.2cm} \caption{{\footnotesize Near
$f_\epsilon\rightarrow{0}$, asimptotic
$\frac{\sqrt{3}f_\epsilon}{4}\varsigma_+ (f_\epsilon)$ from
Eq.~(\ref{sigmas}), which describes fluctuational current above
$T_c$, is not sharply-outlined, but has a smooth decrease. The
fluctuational current is small at high temperatures. }}
\end{center}
\end{figure}
Using the strong field asymptotic $f_\epsilon\gg1,$ we obtain the
fluctuational current at $T_c$
\begin{equation}
j_\infty\equiv
j(f_\epsilon\rightarrow{\infty})=\frac{2}{\sqrt{3}}\frac{eT_c}{\hbar}.
\end{equation}
We wish to point out that $j(f_\epsilon\rightarrow{\infty})/T_c$
is universal and contains only fundamental physical constants. All
material constants like $\xi(0),$ $\lambda(0)$ or the cross
section of the nanowire $S$ are cancelled.
For experimental data processing it is necessary to perform linear
regression of the IV curve
\begin{equation}
\label{universal}
j_\mathrm{tot}
 =\frac{U}{R_N(T_c)}+\mathrm{sign}(U)\times 24.22
 \;\mathrm{nA}\; T_c[\mathrm{K}],
\end{equation}
where the second term is the universal fluctuational current at
strong electric fields around $T_c$. In order to avoid the
thermoelectric effect coming from different materials forming the
contacts to the nanowire one can analyze the current harmonics
predicted by Eq.~(\ref{universal}) for $U(t)=U_0\sin\omega t$ at
$T=T_c$
\begin{equation}
j_\mathrm{tot}(t)=\frac{U_0}{R_N(T_c)}\sin\omega t
 +\frac{4}{\pi}j_\infty\sum_{l=1}^\infty
 \frac{\sin[(2l+1)\omega t]}{2l+1}.
\end{equation}
This current response is a good approximation even slightly above
$T_c$ for voltage amplitudes
\begin{equation}
U_0\gg\frac{8}{\pi}\frac{T_c}{e}\frac{L}{\xi(0)}|\epsilon|^{3/2},
\end{equation}
where $L$ is the length of the nanowire. In such a way the
investigation of fluctuation current in nanowires can be used as a
high accuracy test for applicability of the TDGL equation for
nanostructured superconductors. For further references related to
harmonic generations close to $T_c$ see
Ref.~\onlinecite{Mishonov:02a,Cheenne:03}.

Using Eq.~(\ref{current}) and Eq.~(\ref{distribution}) we derive
1D dimensional density
\begin{equation}
\label{n1DD}
n_{1D}=\frac{n_T}{2\sqrt{\pi}\xi(0)S}{\cal
N}_1(\epsilon,f),
\end{equation}
\begin{figure}
\begin{center}
\includegraphics[width=.55\columnwidth,totalheight=0.25\textheight]{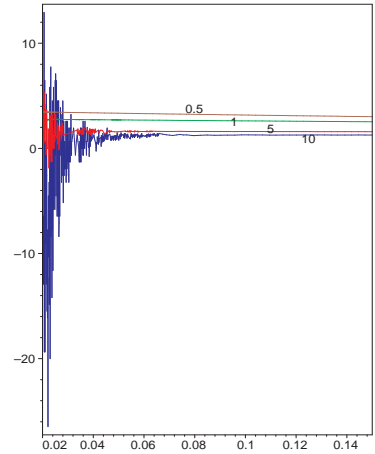}
\vspace*{0.2cm} \caption{{\footnotesize Function ${\cal
N}_1(\epsilon,f)$ from Eq.~(\ref{n1DD}) is at ordinate and
$\epsilon$ - at absciss. The curves illustrate the electric field
$f$ near the phase transition. Very close to the critical
temperature $T_c$ the 1D dimensional density of Cooper pairs has
great fluctuations in strong electric field and small ones in weak
electric field.}}
\end{center}
\end{figure}
\begin{figure}
\begin{center}
\includegraphics[width=.55\columnwidth,totalheight=0.25\textheight]{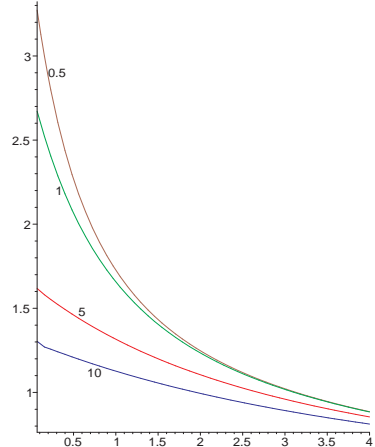}
\vspace*{0.2cm} \caption{{\footnotesize At ordinate is ${\cal
N}_1(\epsilon,f)$ from Eq.~(\ref{n1DD}) and at absciss is
$\epsilon$. The 1D dimensional density decreases with increase the
temperature. The curves illustrate the electric field $f.$ }}

\end{center}
\end{figure}
where ${\cal N}_1(\epsilon,f)$ is previously defined in
Eq.~(\ref{density1D}), when we considered without a derivation the
self-consistent equation for the reduced temperature. The analysis
of temperature and electric field dependence is reduced to two
functions of one variable $f_\epsilon$ above and below the $T_c$
\begin{equation}
n_{1D}=\frac{n_T}{2\xi(0)S\sqrt{|\epsilon|}}N_\pm(f_\epsilon),
\end{equation}
where
\begin{equation}
\label{N_+}
{\cal N}_\pm(f_\epsilon)
 =\frac{1}{\sqrt{\pi}}\int_0^\infty e^{\mp
 v_\epsilon-g_\epsilon v_\epsilon^3}
 \frac{dv_\epsilon}{\sqrt{v_\epsilon}},
 \quad N_+(f_\epsilon=0)=1.
\end{equation}
\begin{figure}
\begin{center}
\includegraphics[width=.55\columnwidth,totalheight=0.25\textheight]{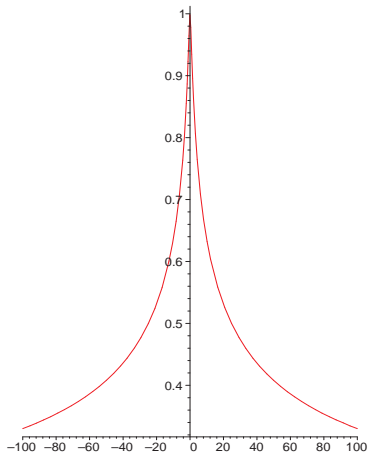}
\vspace*{0.2cm} \caption{{\footnotesize Illustration of ${\cal
N}_+(f_\epsilon)$ from Eq.~(\ref{N_+}) above $T_c$. At absciss is
temperature-electric dependent variable $f_\epsilon$. The density
of Cooper pairs in nanowire is maximum in zero electric field. }}
\end{center}
\end{figure}
\begin{figure}
\begin{center}
\includegraphics[width=.55\columnwidth,totalheight=0.25\textheight]{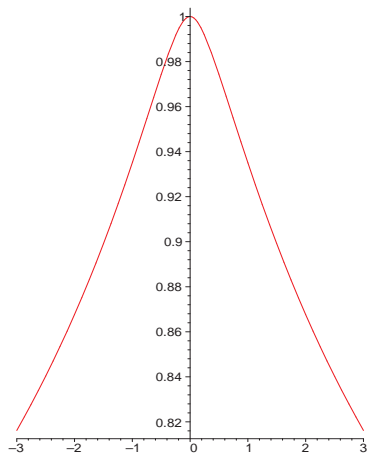}
\vspace*{0.2cm} \caption{{\footnotesize Asimptotic ${\cal
N}_+(f_\epsilon)$ from Eq.~(\ref{N_+}), when
$f_\epsilon\rightarrow{0}$ crosses smoothly ordinate at 1. At
absciss is temperature-electric dependent variable $f_\epsilon$.}}
\end{center}
\end{figure}

Analogously at $T_c$, where $\epsilon=0$ and $f_\epsilon\gg1$ we
obtain
\begin{equation}
n_{1D}(f)=\frac{\Gamma(1/6)n_T}{2^{2/3}3^{5/6}\sqrt{\pi}\xi(0)S}\frac{1}{f^{1/3}}.
\label{3ssiz}
\end{equation}
Choosing a typical set of parameters for the Sn nanowire:
$\xi(0)=1000$~\AA, $\lambda(0)=1000$~\AA, $D=500$~\AA, $S=\pi
D^2/4$ according Eq.~(\ref{Gi1D}) we obtain for 1D Ginzburg number
$\epsilon_{1G}=4.3 \times 10^{-5}.$ This parameter is essential
for the numerical solution of the equation for renormalized
reduced temperature Eq.~(\ref{renormalizedT}). \vspace*{0.5cm}

\section{Discussion and Conclusion}

Solving in parallel the TDGL equation and the Boltzmann equation
we obtained coinciding results: not only for the linear case of
Aslamazov-Larkin conductivity, but for the cases of strong
electric fields, arbitrary time dependence of the electric field,
nonparabolic momentum dependence of energy of Cooper pairs, energy
cut-off, self-consistent equation for the renormalized reduced
temperature, frequency dependence of the fluctuation conductivity
etc. The number of fluctuation Cooper pairs which participates in
the Boltzmann equation and the formulas for the current is
actually the diagonal element of the order parameter
correlator\cite{Silva:03}
$n_k(t)=C\left[p_\mathrm{kin}=p-e^*A(t);t,t\right].$ One can also
easily check that the entropy of fluctuation Cooper pairs $\eta$
is increasing with the time $d\eta/dt\geq0;$ the capital $\eta$ in
the $\eta$-theorem by Boltzmann is often spelled as Latin $H.$ Our
self-consistent formula for the fluctuation conductivity of a
superconducting nanowire can be directly used for the experimental
data processing. In such a way we conclude that Boltzmann equation
for fluctuation Cooper pairs reproduces the known results of the
fluctuation theory in the normal phase and it is a adequate tool
to predict new phenomena related to metastable Cooper pairs, like
negative differential conductivity in the fluctuation regime
predicted in Ref.~\onlinecite{Mishonov02} and strong electric
field effect in nanostructured superconductors where the heating
effects are reduced. Our universal result Eq.~(\ref{universal})
for a fluctuational current in a nanowire under strong electric
field shows that Boltzmann equation for the fluctuation Cooper
pairs will become an indispensable tool for understanding the
electronic processes in nanostructured superconductors.

\acknowledgments Two of the authors T.~M. and D.~D., would like to
thank to  M.~Ausloos and A.~Varlamov for the hospitality in NATO
ASI in Trieste where the present results for derivation of the
Boltzmann equation, (cf. Ref.~\onlinecite{Mishonov-Damianov}) have
been presented. The authors are thankful to J.~O.~Indekeu for the
hospitality in K.~U.~Leuven where the present work was completed.
One of the authors, T.~M., is thankful to E.~Abrahams, M.~Ausloos,
and L.~P.~Gor'kov for the correspondence related to his papers and
appreciates the discussions on the Boltzmann kinetic equation with
J.~O.~Indekeu, A.~I.~Larkin, S.~Michotte, E.~S.~Penev,
L.~P.~Pitaevskii, A~Rigamonti, E.~Silva, and A.~Varlamov. He is
very much indebted to late L.~G.~Aslamazov for introducing him
long ago in the problem of fluctuation conductivity. This research
has been partially supported by Flemish program GOA.


\end{document}